\documentclass[amsfonts,aps,prl,twocolumn]{revtex4}
\usepackage{graphicx}

\newcommand{\placement}{p}
\newcommand{\mb}[1]{\mathbf{#1}}
\newcommand{\ind}[1]{{\textrm{\tiny{#1}}}}
\newcommand{\eww}[1]{\la{#1}\ra}

\newcommand{\equ}[1]{Eq.~(\ref{#1})}

\newcommand{\figref}[1]{Fig.~\ref{#1}}

\newcommand{\tauPhi}{\eww{\tau}}
\newcommand{\tauPhiT}{\eww{\tau(T)}}

\renewcommand{\d}{{\rm d}}

\newcommand{\xmin}{\xi}
\newcommand{\eps}{\epsilon}

\newcommand{\pos}{\mb{x}}
\newcommand{\la}{\left\langle}
\newcommand{\ra}{\right\rangle}
\renewcommand{\phi}{\varphi}
\newcommand{\kB}{k_{\rm B}}
\newcommand{\NEQU}[2]{\begin{equation}#1\label{#2}\end{equation}}
\newcommand{\EQU}[2]{$$#1$$}

\begin{document}
\title{Hopping in a Supercooled Lennard-Jones Liquid:
Metabasins, Waiting Time Distribution, and Diffusion}
\author{B. Doliwa$^1$ and
A. Heuer$^2$}
\affiliation{$^1$Max Planck Institute for Polymer Research, Mainz, Germany}
\affiliation{$^2$Institute of Physical Chemistry, University of M\"unster -
  M\"unster, Germany}
\date{\today}
\begin{abstract}
We investigate the jump motion among potential energy minima of a
Lennard-Jones model glass former by extensive computer
simulation. From the time series of minima energies, it becomes
clear that the energy landscape is organized in superstructures,
called metabasins.  We show that diffusion can be pictured as a
random walk among metabasins, and that the whole temperature
dependence resides in the distribution of waiting times. The
waiting time distribution exhibits algebraic decays:
$\tau^{-1/2}$ for very short times and
$\tau^{-\alpha}$ for longer times, where $\alpha\approx2$ near $T_c$.
We demonstrate that solely the waiting
times in the very stable basins account for the temperature dependence
of the diffusion constant.
\end{abstract}
\maketitle
\newcommand{\xs}{x_\ind{s}}
\newcommand{\xmb}{\xi_\ind{MB}}
\newcommand{\pback}{p_\ind{back}}
\newcommand{\taustar}{\tau^*}
\newcommand{\taupsi}{\eww{\tau}_\psi}
\newcommand{\Eact}{E_\ind{act}}
\vspace{2cm}
The energy landscape picture proposed more than thirty years ago
by Goldstein~\cite{Goldstein:1969} has turned out to be a fruitful
way of describing the complicated many-particle effects in
disordered systems~\cite{Angell:332}.  Starting from the joint
potential energy landscape (PEL) $V(x)$ of $N$ particles as a
function of their configuration $x=\{\pos_1,...\pos_N\}$ one
expects that the properties of the system at sufficiently low
temperatures will be dominated by long residences near local
minima of $V(x)$ ({\it inherent structures}) with rare hopping
events between them~\cite{Stillinger:333}. Recently it became clear
for a model glass former that the strict hopping picture
approximately holds for $T < T_c$ ({\it landscape-dominated
regime})~\cite{Schroder:88,Angelani:212,Broderix:228} where $T_c$ is the
mode-coupling temperature~\cite{Gotze:1992}. However, even for higher
temperatures $T < 2 T_c$ many dynamic properties are still related
to the properties of inherent structures ({\it
landscape-influenced regime})~\cite{Sastry:209,Scala:71,Buchner:11}.
Thus, in both temperature intervals an observable like the
diffusion constant $D(T)$ should depend on
the topography of inherent structures (IS) or, more generally, of
basins (a basin of an inherent structure is defined as the set of
configurations that reach this minimum via steepest descent~\cite{Stillinger:222}).
Such an understanding is indispensable to
grasp the underlying physics of the Adam-Gibbs relation~\cite{Adam:39,Richert:218}.

A simplified picture of glassy dynamics has been expressed in
phenomenological models
in~\cite{Zwanzig:1983,Monthus:310,Odagaki:328} based on spatially
uncorrelated hopping processes ({\it random-walk}) in
configuration space.
Then the whole temperature dependence is contained in
the average waiting time $\tauPhiT$.
The true dynamics of glass forming
systems, however, is expected to be more complicated.
For example it is known
that back- and forth correlations cannot be
neglected and that the elementary jump distances
depend on temperature~\cite{Schulz:338,Gaukel}.
In general, in a hopping approach the temperature dependence of
the diffusion constant may be related to spatial and temporal
aspects, as expressed by the relation
\NEQU{D(T) = \frac{a^2(T)}{6N\tauPhiT}.}{EQU1}
With this ansatz,
we anticipate
the important role of the mean waiting time
and collect the spatial details of hopping in
an {\it effective} jump width $a(T)$.
The latter involves
(i)~the average jump distance,
(ii)~correlations of jump widths with waiting times,
and
(iii)~directional correlations of successive jumps.
To our knowledge this decomposition into spatial and temporal
contributions has not been systematically implemented
within the PEL framework so far.
A priori it is not clear to which degree the temperature dependence
of $a(T)$ is relevant; see, e.g.,~\cite{Vardeman:339,Schulz:321}.
Some information about the waiting time distribution~(WTD)
has already been gained from the analysis of hopping
processes of single particles in real space via computer
simulations~\cite{Miyagawa:69,Allegrini:1}.
In contrast, we consider hopping in configuration space, with the advantage
of incorporating the full many-particle effects~\cite{Keyes:335}.

In this paper, we present detailed information about the spatial
and temporal aspects of hopping in a model glass former,
and individually determine $a(T)$ and $\tauPhiT$.
We demonstrate that (i)~{\it only}
$\tauPhiT$ depends on temperature,
(ii)~hopping among single
basins is not a random walk, whereas hopping among superstructures
of minima (metabasins) is close to a random walk,
(iii)~$\tauPhiT$ is dominated by the long waiting times, due to
a slow (approximately algebraic) decay of WTDs.

In the present work, we investigate a binary mixture of
Lennard-Jones particles (BMLJ), as recently
treated by two groups~\cite{Broderix:228,Wales:286};
see also~\cite{Kob:203}.
It is characterized by the interaction
potentials
$V_{\alpha\beta}(r)=4\epsilon_{\alpha\beta}[(\sigma_{\alpha\beta}/r)^{12}-(\sigma_{\alpha\beta}/r)^{6}]$
with the parameter set $N=N_A+N_B=52+13=65$,
$\sigma_{AB}=0.8\sigma_{AA}$, $\sigma_{BB}=0.88\sigma_{AA}$,
$\epsilon_{AB}=1.5\epsilon_{AA}$, $\epsilon_{BB}=0.5\epsilon_{AA}$, $r_c=1.8\sigma_{AA}$.
Linear
functions were added to the potentials to ensure continuous forces
and energies at the cutoff $r_c$.
Units of length, mass, energy, and time are $\sigma_{AA}$, $m$, $\epsilon_{AA}$,
and $\sqrt{m\sigma_{AA}^2/\eps_{AA}}$, respectively.
For convenience, though, we will omit units here.
We use Langevin molecular
dynamics simulations (MD) with fixed step size,
$\lambda^2=0.015^2=2\kB T\Delta t/m\zeta$,
equal particle masses $m$,
friction constant $\zeta$,
and periodic boundary conditions at a density of $\rho=1.2$.
The friction constant $\zeta=2/0.015^2$ is chosen so that
$\Delta t=1/T$.
Due to the different type of dynamics, the {\it absolute} values
of times and diffusion constants are different from
those found within Newtonian dynamics simulations.
The mode-coupling temperature is
$T_c=0.45\pm0.01$ in this model system (compare~\cite{Kob:203}).

For the analysis of
dynamics in configuration space it is essential to use small
systems because otherwise many interesting effects will be
averaged out~\cite{Buchner:11}.
The relevance of small systems has been also
pointed out by other groups; see, e.g.~\cite{Keyes:335,Grigera:264}.
On the other hand, the system
should not be too small in order to avoid major finite size effects.
$N \approx 60$ turns out to be a very good compromise for binary
Lennard-Jones mixtures, whereas $N\le 40$ already displays major finite-size
effects~\cite{Buchner:193}. Here we choose $N=65$.
To back those findings,we have carried out an extensive study of finite-size effects for systems
of $N=65,130$, and $N=1000$ particles~\cite{Doliwa:404}.
It turned out that the $N=65$ system is nearly identical to the bulk ($N=1000$)
above $T_c$. Since well-equilibrated runs of $N\ge130$
are lacking below $T_c$, finite-size effects cannot be excluded there at the present stage.
However, this does not affect the main results of this paper.
More important is the question of good equilibration at each temperature.
Have the runs been long enough to sample the PEL sufficiently?
Above $T_c$ this is uncritical, which can be seen from the fact
that each run comprised at least 850 $\alpha$-relaxation times.
A more detailed check, involving the lifetimes and distribution of metabasins,
indicates that runs down to $T=0.435$ are feasible with the available computer power.


By regular quenching the MD trajectory $x(t)$ to the bottom of the
basins visited at time $t$, as proposed by Stillinger and Weber,
we obtain a discontinuous trajectory $\xmin(t)$. In this way, one
discards the more or less complicated vibrational part
$x(t)-\xmin(t)$ of motion, only keeping the visited minima as
'milestones'.  The one-particle diffusion constant can be also
determined from the
squared displacement of inherent
structures via
$D=\lim_{t\to\infty}\eww{(\xmin(t)-\xmin(0))^2}/6Nt$.

How to resolve the {\it elementary} hopping events?
Since computer time prohibits to
calculate $\xmin(t)$ for every time step, we
normally find ourselves in the situation of having equidistant
quenched configurations $\xmin(t_i)$, with, say,
$t_{i+1}-t_i\sim10^5$~MD~steps.
If the same minimum is found for
times $t_i$ and $t_{j}$, we need not care about transitions in the
meantime, because no relaxation has occurred there.  If, in
contrast, $\xmin(t_i)\ne\xmin(t_{i+1})$, we must not expect
$\xmin(t_{i+1})$ to be the direct successor of $\xmin(t_{i})$,
since many other minima could have been visited between $t_i$ and
$t_{i+1}$.  Therefore, further minimizations in this time interval
are necessary. For reasons of efficiency, we apply
a straightforward interval bisection method, which locates
all
relevant transitions with an accuracy of 1~MD~step. Although
computationally demanding,
this
has proven most
efficient for resolving the relevant details of hopping on the
PEL.

\begin{figure}[\placement]
\begin{center}
\includegraphics[width=7cm]{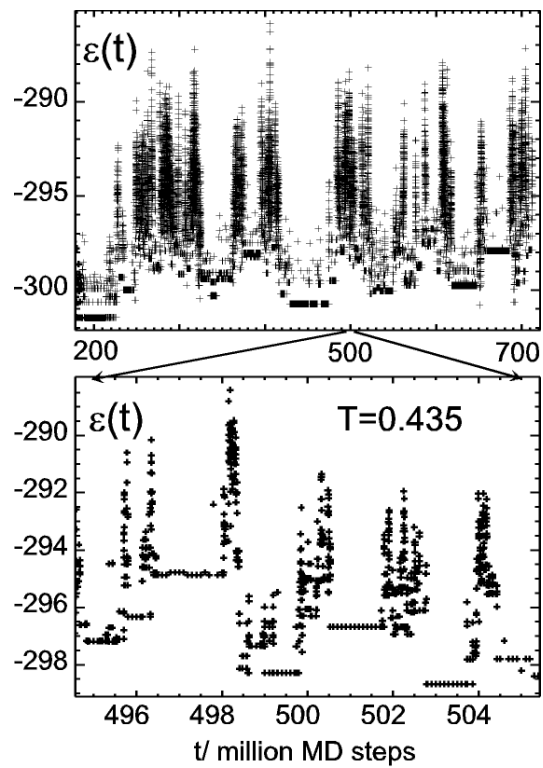}
\end{center}
\caption{ The time series of minima energies measured for the BMLJ
system of $N=65$ particles, $T=0.435$. The distance between
minimizations before interval bisection is $10^5$~MD~Steps.  The
length of the total run is $2\times10^9$~MD~Steps. Top: time
window covering a quarter of the total run. Bottom: magnification by a
factor of fifty.  } \label{FIG1}
\end{figure}
As demonstrated in~\cite{Buchner:11}, the time series of potential
energies $\eps(t)=V(\xmin(t))$ reflects well the character of dynamics
in the supercooled state.
For $T=0.435$, $\eps(t)$ is shown in \figref{FIG1}
from which we note a remarkable structure in $\eps(t)$.
The system is trapped in some stable
configurations
for long times, during which a small number of minima is visited
over and over again. Obviously, these minima form superstructures,
which,
following~\cite{Stillinger:333},
we denote {\it metabasins} (MBs).
One may imagine that minima of long-lived MBs are organized
in funnel-like structures so that the system
is stuck there for a long time.
It has been argued that the occurrence of $\beta$-relaxation
at low temperatures is due to such substructure of the PEL~\cite{Stillinger:333}.
This is supported by the real-space signature of MBs as reported
by Middleton and Wales~\cite{Middleton:214}.
Formally there is no
unique way to define MBs for a given PEL due to the lack of
a strict time scale separation. Here we take the pragmatic view
and let the system decide by its MD run.
The intuitive notion of MBs from \figref{FIG1} can be cast into
an algorithm in a straightforward way, see~\cite{Buchner:11}~\footnote{
It is important to note that, different
from~\cite{Buchner:11}, we will treat all MBs on the same footing
here, no matter if they contain one or many minima.
}.
One major advantage of analyzing MBs rather
than basins is that the charming simplistic picture of a random
walk in configuration space will be better fulfilled
since direct back- and forth
correlations are already taken into account. It turns out (see
\figref{FIG1}) that long-lived stable MBs are separated by
bursts of rapid transitions among higher minima which look like
fountains.
They cannot be detected without interval bisection. The
waiting times of the MBs range from a few MD~steps to
many millions of them. Moreover, comparing \figref{FIG1} (a) and
(b), we find a certain self-similarity in $\eps(t)$ when inspected
on different time scales.
The distribution of MB lifetimes $\tau$
will be denoted $\phi(\tau,T)$, its first moment $\tauPhiT$ being a key
quantity for the
following considerations.

\begin{figure}[\placement]
\begin{center}
\includegraphics[width=7.5cm]{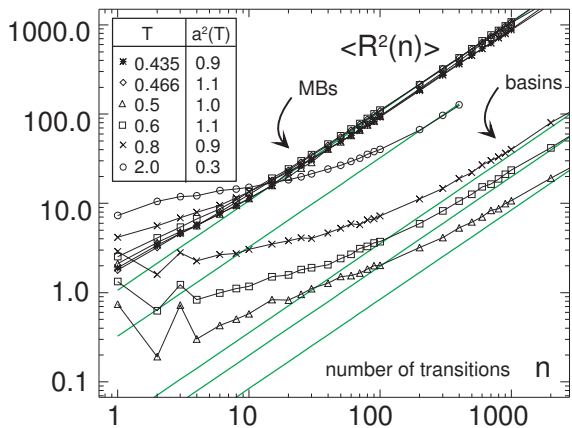}
\end{center}
\caption{
Squared displacement
$\eww{R^2(n)}$
after $n$ MB jumps for the temperatures
$T=0.435,0.466,0.5,0.6,0.8$, and~$2.0$.
Also shown are three curves ($T=0.5$, $0.6$, and $0.8$)
for jumps among single basins (lower curves). The latter data have
been obtained by extremely frequent minimizations, which resolve
nearly all IS transitions.
We have included lines of slope 1.
} \label{FIG2}
\end{figure}
Our goal is to find an expression for the effective jump width $a(T)$
of \equ{EQU1}.
To this end, as a generalization of $\xmin(t)$, we
define the MB inherent structure $\xmb(t)$
as the mean of $\xi(t)$ over the MB lifetime.
The key idea now is to introduce the squared
distance $\eww{R^2(n)}$ after $n$ jumps~\cite{Maass},
averaged over different
realizations. This quantity is purely spatial since it does not
involve any time scale. In the limit of
large $n$
one obtains
due to the central limit theorem $ \lim_{n \rightarrow \infty}
\langle R^2(n)\rangle /\langle \xmb^2(n \tauPhi)\rangle = 1$ where
$\langle \xmb^2(n \tauPhi)\rangle$ is the average squared
displacement after time $n\tauPhi$.
Thus,
\EQU{D(T) =
\lim_{n\to\infty}\frac{\eww{\xmb^2(n\tauPhi)}}{6Nn\tauPhi}
=\left [
\lim_{n\to\infty}\frac{\eww{R^2(n)}}{6n}\right]\frac{1}{N\tauPhi}}{}
where the first factor may be identified as $a^2(T)/6$.
Note that both factors are independent of system size because
IS transitions are localized ($\eww{R^2(1)}=O(1)$) and
mean waiting times decrease with system size
($\tauPhi=O(1/N)$).
We may now calculate $a(T)$ from the simulations,
results are shown in~\figref{FIG2}.
The most important observation is that $a(T)$ is
temperature independent
for $T<1$.
Interestingly, this is not affected by the variation of the
elementary jump width $\eww{R^2(1)}$, which
increases with temperature, a fact that was recently observed
by Schulz~et~al.~\cite{Schulz:338}.
A possible explanation for this might be found in the
increasing population of
higher-order stationary points~\cite{Angelani:212}, which
are known to be more distant to neighboring minima,
but also provoke  many 'bookkeeping' IS transitions, thus
resulting in larger backward correlations~\cite{Keyes:335}.

In any event, the constancy of $a(T)$ in the landscape-influenced
regime $T<1$
implies that the temperature dependence of $D(T)$ follows alone
from $\tauPhiT$,
i.e. $D(T)\propto1/\tauPhiT$.
This simple picture
breaks down for $T>1$ where
the explored regions of configuration space
probably have a completely different structure.
It has to be noted that the constancy of $a(T)$
for low T
relies heavily on
our resolution of {\it all} elementary IS transitions leading to
relaxation.

A further insight from \figref{FIG2} is that the dynamics on the level
of MBs is basically a random walk except for minor back-correlations
for $n\le 5$.  As seen from the figure, these correlations are
present between {\it single} basins, the consequence being a
significant deviation from the relation $\eww{R^2(n)} \propto n$.
Also note the oscillations in the single-basin $\eww{R^2(n)}$ at small
$n$, which result from the back-and-forth motion within MBs.  More
importantly, the single-basin curves do not have the same large-$n$
limit so that the effective jump length on the level of single basins
would be temperature {\it dependent}.
It remains unclear why for a very small
LJ system ($N=32$) correlations among adjacent basins are
irrelevant for $T \approx T_c$~\cite{Keyes:335}, and why intra- rather
than inter-basin dynamics is deemed to be the key to the understanding
of diffusion~\cite{Keyes:262}.

\begin{figure}[\placement]
\begin{center}
\includegraphics[width=6.5cm]{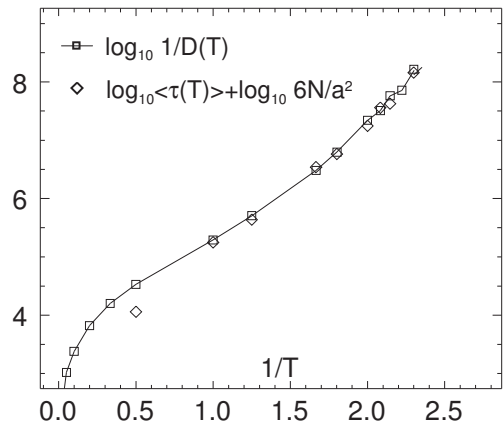}
\end{center}
\caption{ Arrhenius plot of the inverse one-particle diffusion
constant $1/D(T)$ and the mean waiting time
$\tauPhiT$
multiplied by a constant ($a^2=1.0$).
Error bars are of the order of the symbol size.
} \label{FIG3}
\end{figure}
We can check the relation
$D \propto 1/\tauPhi$
within our
simulations.
Figure~3 shows that it is indeed well
fulfilled for $T<1$ while for $T=2$ we find the expected deviations.

In the remaining part of the paper, we discuss the properties of
the WTDs $\phi(\tau,T)$, see Figure~4.
For short $\tau$, all curves exhibit a power-law behavior with exponent $-1/2$,
similarly to~\cite{Leporini:336}.
At $\tau\approx10^3 - 10^4$ a crossover to
the faster decay
$\phi(\tau,T)\propto\tau^{-\alpha(T)}$
can be observed.
For $T \approx 0.45$, one finds
$\alpha\approx2.0$ for which the expectation value $\tauPhi$ would
diverge. However, the behavior $\phi(\tau)\propto\tau^{-\alpha}$
cannot extend to infinity.
Due to the finite number of MBs in the system,
there
exists a maximum effective barrier $E_\ind{max}$, giving rise to an exponential
cutoff at some minimum rate $\gamma_\ind{min}$.
\begin{figure}[\placement]
\begin{center}
\includegraphics[width=7cm]{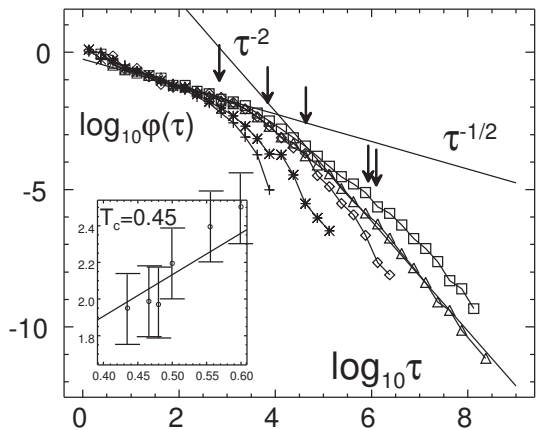}
\end{center}
\caption{
Distributions of waiting times, $\phi(\tau,T)$,
for $T=1.0,0.8,0.6,0.5$, and $0.435$ (from left to right).
Curves have been shifted to overlap for small $\tau$.
Lines corresponding to algebraic decays with exponents
$0.5$ and $2$ are shown as guides to the eye.
The arrows mark $\tau^*(T)$, i.e. the waiting time with the
property $\int_{\tau^*(T)}^\infty\d\tau\phi(\tau)\tau =0.9\tauPhi$.
Inset: Power-law exponent $\alpha(T)$
from fits to the long-time decay.
Within the possible accuracy, $\alpha(T)\approx T/T_c+1$.
} \label{FIG4}
\end{figure}

The slow decay of the WTDs leads to the important observation that
the mean waiting time $\tauPhi$ is dominated by the contributions
from large $\tau$ (see arrows in figure~4). For example at
$T=0.435$, ninety percent of $\tauPhi$ are made up by the
six percent longest waiting times,
which are basins of lifetimes
greater than $0.5\times10^6$~MD~Steps.
This may come as a surprise because
(i) the short-lived MBs are much more numerous
than the long-lived
and (ii) one might intuitively think
that the diffusion constant and thus
$\tauPhi$ is dominated by the fast particles.
The result
is in qualitative agreement with the approach of Wolynes and Xia
who regard the relaxation of long-lived local structures as the
time-determining step~\cite{Xia:233}.

Interestingly, the algebraic decay of the WTDs follows for some
theoretical models of diffusion with built-in traps.
This is the case for Bouchaud's trap model~\cite{Monthus:310} and
the trapping diffusion model of Odagaki~et~al.~\cite{Odagaki:328}.
A recent comparison of WTDs in the Lennard-Jones system with that of
trap models can be found in~\cite{Reichman:398}.

In conclusion, the detailed analysis of hopping on the PEL has
provided new insights into the mechanism of diffusion in
supercooled liquids. As we have seen, the emergence of long-lived
MBs is the reason for the slowing down of molecular motion
in our supercooled model liquid.
As a next step the waiting time
should be related to the respective MB energies to
establish a connection between energy (thermodynamics) and
dynamics in the spirit of the Adam-Gibbs relation
(see the subsequent paper~\cite{doliwa:392}).

We gratefully acknowledge helpful discussions with J.P.~Bouchaud,
T.~Odagaki, D.R.~Reichman, R.~Schilling, H.R.~Schober, and H.W.~Spiess. Funding
was granted by the Sonderforschungsbereich 262.

\bibliographystyle{apsrev}
\bibliography{phi}

\end{document}